\documentclass[12pt]{article}
\usepackage{axodraw,epsfig,psfrag}
\def\be{\begin{equation}}
\def\ee{\end{equation}}
\def\bea{\begin{eqnarray}}
\def\eea{\end{eqnarray}}
\def\nn{\nonumber}
\def\st{\tilde{t}}

\def\stl{\tilde{t}_L}
\def\str{\tilde{t}_R}
\def\scl{\tilde{c}_L}
\def\scr{\tilde{c}_R}
\def\sq{\tilde{q}}
\def\glu{\tilde{g}}
\def\mt{\tilde{m}}
\def\mmt{\tilde{m}^2_{t}}
\def\mmc{\tilde{m}^2_c}
\def\mmtl{\tilde{m}^2_{t_L}}
\def\mmtr{\tilde{m}^2_{t_R}}
\def\mmcl{\tilde{m}^2_{c_L}}
\def\mmcr{\tilde{m}^2_{c_R}}
\def\mmul{\tilde{m}^2_{u_L}}
\def\mmel{\tilde{m}^2_{e_L}}
\def\mmmul{\tilde{m}^2_{\mu_L}}
\def\mmtaul{\tilde{m}^2_{\tau_L}}
\def\tca{A_{tc}}
\def\cta{A_{ct}}
\def\xxt{|X_t|^2}
\def\aatc{|\tca|^2}
\def\aact{|\cta|^2}
\def\xt4{|X_t|^4}
\def\atc4{|\tca|^4}
\def\act4{|\cta|^4}
\def\de{\delta}

\def\ga{\gamma}
\def\la{\lambda}

\def\ov{\overline}   
\def\simlt{\stackrel{<}{{}_\sim}}
\def\simgt{\stackrel{>}{{}_\sim}}
\def\to{\rightarrow}
\begin{document}
\begin{titlepage}
\phantom{nothing}
\vskip 2.0cm
\begin{center}
{\Large\bf 
The supersymmetric Higgs boson
\\ 
\vskip 0.5 cm
with flavoured $A$-terms} 
\end{center}
\vskip 1.5  cm
\begin{center}
{ 
{\large Andrea Brignole}
\vskip .5cm
{\it INFN, Sezione di Padova, via Marzolo 8, I-35131 Padova, Italy}
}
\end{center}
\vskip 3.0cm
\begin{center}
{\bf Abstract}
\end{center}

We consider a supersymmetric scenario with large flavour violating
$A$-terms in the stop/scharm sector and study their impact 
on the Higgs mass, the electroweak $\rho$ parameter
and the effective Higgs couplings to gluons, photons and 
charm quarks. For each observable we present explicit analytical 
expressions which exhibit the relevant parametric dependences,
both in the general case and in specific limits. We find
significant effects and comment on phenomenological 
implications for the LHC and future colliders.
\end{titlepage}
\setcounter{footnote}{0}


\section{Introduction}

The recent discovery of a Higgs particle at the LHC 
\cite{higgsdisc} has confirmed the
validity of the Standard Model (SM). ATLAS and CMS have 
found that the Higgs boson has a mass of 125~GeV \cite{mass15}
and that its couplings to gauge bosons and third-generation fermions 
are consistent with the SM predictions \cite{cms14,atlas15}. 
The experimental uncertainties on Higgs
couplings are still sizeable, but they will be progressively
reduced, first at the LHC and then at future colliders.
Several models of physics beyond the SM will be tested
as well, since they generically predict both deviations 
in those couplings and new particles to be discovered. 
Supersymmetric (SUSY) extensions of the SM are well known
examples of such theories. In a wide class of SUSY models,
including the MSSM and the NMSSM, the Higgs sector contains 
two doublets and possibly singlets \cite{mssm,djou,nmssm}. 
A linear combination of the doublets 
($H = \sin\beta \, H_u - \cos\beta \, i \sigma_2 H_d^*$)
behaves as the SM doublet in the decoupling limit of the other
Higgs states. We will focus on this regime for definiteness.
Our aim is to investigate some properties of such SM-like Higgs field 
in a particular region of the SUSY parameter space,
characterized by large values of certain flavour violating
trilinear couplings. 

Let us parametrize the neutral component of $H$ as 
$H^0=v+\frac{1}{\sqrt{2}}(h+iG)$, 
where $v\simeq 174 \, {\rm GeV}$ triggers
$SU(2)\times U(1)$ breaking, $h$ is the physical Higgs boson
and $G$ is the neutral would-be Goldstone boson.
Stop squarks are the SUSY particles that couple more
strongly to $H^0$. Indeed, $F$-terms generate large quartic 
couplings of the form $y_t^2 |H^0|^2  (|\stl|^2 + |\str|^2)$,
where $y_t$ is the SM top Yukawa coupling ({\em i.e.},  $y_t$ is related 
to the top mass through $m_t = y_t \, v$ at the tree level). 
$F$-terms and SUSY-breaking $A$-terms also generate
Higgs-stop-stop trilinear couplings. We assume 
that sizeable Higgs-stop-scharm $A$-terms are present as well,
although we do not specify their origin. 
Hence the trilinear scalar interactions that are relevant 
to us are
\be
\label{vtril}
V_{(3)}  =   
y_t \, H^0 \, (X_t \, \str^{\,*} \, \stl 
+ \tca \, \str^{\,*} \, \scl + \cta \, \scr^{\,*} \, \stl)
+ {\rm h.c.} \; ,
\ee
where we have used the standard notation 
$X_t \equiv A_t - \mu^* \cot \beta$ and factored out $y_t$ 
for convenience. We complete our parametrization by
writing stop and scharm (SUSY-breaking) mass terms as
$\mmtl |\stl|^2 + \mmtr |\str|^2 + \mmcl |\scl|^2 + \mmcr |\scr|^2$.
Gauge invariance implies mass terms 
$ \mmtl |\tilde{b}_L|^2 + \mmcl |\tilde{s}_L|^2$ for $\tilde{b}_L$ and
$\tilde{s}_L$, the $SU(2)$ partners of $\tilde{t}_L$ and $\tilde{c}_L$. 
In general one also expects flavour violating mass terms 
of $LL$ and $RR$ type, as well as other $LR$ trilinears. 
The effective low-energy values of all such parameters depend, as usual, 
both on boundary conditions at some higher scale and on renormalization 
effects, which include those generated by $\tca$ and $\cta$ themselves. 
Although we are aware of the latter connection, we decide to explore 
the region of the phenomenological SUSY parameter space where all
flavour violating masses are small, apart from those in eq.~(\ref{vtril}).
In other words, we do not specify either a flavour model or a mechanism
of SUSY breaking, treat $\tca$ and $\cta$ as phenomenological parameters
(on the same footing as $X_t$) and allow them to be as large as the 
flavour conserving stop and scharm masses. 
Even in the latter limit the constraints from flavour changing 
observables are weak or absent, especially for ${\cal O}({\rm TeV})$
squark masses.
We will return to this point in a separate paper \cite{bp}, whereas here
we will study the impact of $\tca$ and $\cta$ on a set of
{\em flavour conserving} quantities, namely:
\begin{description}
\item[{\em i)}] the mass of the Higgs boson $h$ (Section~2); 
\item[{\em ii)}] the $\rho$ parameter (Section~3); 
\item[{\em iii)}] the effective coupling of $h$ to gluons 
or photons (Section~4); 
\item[{\em iv)}] the effective coupling of $h$ to charm quarks 
(Section~5). 
\end{description}
In each case we will compute the leading effects and present
simple analytical expressions which exhibit the relevant
parametric dependences. By `leading' we mean that
squarks are integrated out at the one-loop level,
at leading order in $y_t$ and at lowest order in $v^2/\mt^2$,
where $\mt^2$ generically denotes a squark mass. In the effective theory
language, the latter point means that we evaluate the squark
contribution to the operators of lowest dimension ($d$) associated 
with each of the above quantities, namely:
\begin{description}
\item[{\em i)}] $|H^0|^4$ ($d=4$);  
\item[{\em ii)}] $|H^{\dagger} D_{\mu} H|^2$ ($d=6$); 
\item[{\em iii)}] $|H^0|^2 G_{\mu\nu} G^{\mu\nu} \, , \, 
|H^0|^2 F_{\mu\nu} F^{\mu\nu}$ ($d=6$);
\item[{\em iv)}] $|H^0|^2 H^0 \, \ov{c_R}\, c_L+ {\rm h.c.} $ ($d=6$).
\end{description}
These different dimensionalities imply different decoupling
properties, of course. Squark contributions 
to $\rho$ and to the effective Higgs couplings are suppressed 
by a factor $v^2/\mt^2$ ($d=6$ operators). On the other hand,
corrections to the Higgs mass do not feel that suppression
since they are associated to a quartic coupling ($d=4$ operator).
In any case, once the appropriate power of $v^2/\mt^2$ is taken
into account, for each of the above quantities the remaining
dependence on mass parameters will be encoded in some dimensionless 
function.
In particular, the dependence on trilinear parameters will appear 
through powers of $X_t/\mt,\tca/\mt,\cta/\mt$.   


\section{The Higgs mass}

In the SM the Higgs mass is $m_h^2 = \la v^2$ at the tree level, 
where $\la$ is a free parameter that controls the quartic 
term of the Higgs potential ($V \supset \frac{1}{4}\la |H^0|^4$).
The measured value of $m_h \simeq  125$~GeV implies
$\la \simeq 0.5$ at the weak scale.
In SUSY scenarios with a SM-like Higgs, $\la$ is an effective coupling 
that can receive contributions from different sources: 
$\la \simeq \sum_i \de \la_i$. 
At the tree level the standard  $D$-term contribution 
$\de \la_D =\frac{1}{2}(g^2+g'^2) \cos^2 2\beta$ predicts
$m_h = m_Z |\cos 2\beta| \leq m_Z$, significantly lower than
125~GeV. Additional tree-level contributions 
to $\la$ can arise, {\it e.g.}, from $F$-terms
in extensions with singlets \cite{nmssm}
or from higher dimension effective operators 
\cite{higher}. Important contributions to $\la$
also arise radiatively. The leading ones are generated by
top and stop one-loop diagrams \cite{mh90} and are proportional
to $y_t^4$:
\be
\label{dela}
\de \la_{\rm log} \simeq \frac{3 y_t^4}{4 \pi^2} 
\log \frac{\mmt}{m_t^2} 
\;\;\; , \;\;\; 
\de \la_{\rm thr} \simeq \frac{3 y_t^4}{4 \pi^2} \, \Delta  \, .
\ee
The first term $\de \la_{\rm log}$ can be interpreted either as
the combination of logarithmically divergent top and stop
contributions, or as the result of the 
top-loop-induced running of $\la$ from the stop mass scale 
$\mmt \simeq (\mmtl \mmtr)^{1/2}$ to the weak scale
($4 \pi^2 d\la/d \log q^2 =-3 y_t^4+\ldots$). The second term
$\de \la_{\rm thr}$ is a finite threshold correction
at the stop scale and we parametrize it through $\Delta$, 
a dimensionless function of squark mass
parameters\footnote{More precisely, the tree-level contributions
and $\de\la_{\rm thr}$ determine the SM coupling $\la$ 
at the SUSY matching scale. The latter coupling is then 
renormalized down to the weak scale
where $m_h$ is evaluated. For recent investigations on  
the SUSY threshold effect, RG evolution and higher order
corrections, see \cite{bgss} 
and refs. therein. If one uses the approximate expressions 
in eq.~(\ref{dela}), the factors $y_t^4$ may be 
evaluated at the scales suggested in \cite{hhh}, {\em i.e.}
at $(m_t\mt_t)^{1/2}$ in  $\de \la_{\rm log}$ and at $\mt_t$ in
$\de\la_{\rm thr}$.}. In the flavour conserving limit,
$\Delta$ contains quadratic and quartic powers of
$X_t$. As well known, an appropriate
choice of $X_t$ can give a substantial contribution
to $\Delta$, which translates into a correction
$\de m_h^2 \simeq \de \la_{\rm thr} v^2$ to the Higgs 
mass. In the often quoted limit
$\mmtl=\mmtr=\mmt$, for instance, 
$\Delta=|X_t|^2/\mmt - \frac{1}{12}|X_t|^4/\mt^4_t$ reaches 
its maximal value $\Delta_{\rm max}=3$ at $|X_t|= \sqrt{6} \,\mt_t$.
The corresponding linear correction to $m_h$ is about $15$--$20$~GeV.
We recall that, as often emphasized,
a large threshold correction is welcome because it allows a smaller
stop mass scale in the logarithmic term 
({\em e.g.}, $\mt_t$ around  1~TeV rather than in the multi-TeV range).
However, the knowledge of the threshold correction is important also 
in more general scenarios.

\begin{figure}[tb]
\begin{center}
\begin{picture}(100,70)(-50,-35)
\DashCArc(0,0)(15,0,360){3}
\DashLine(0,15)(-10,25){2}
\DashLine(0,15)(10,25){2}
\DashLine(-10.5,-10.5)(-20,-20){2}
\DashLine(10.5,-10.5)(20,-20){2}
\Text(-22,5)[]{\small $\stl$}
\Text(22,5)[]{\small $\stl$}
\Text(2,-23)[]{\small $\str$}
\Text(-15,30)[]{\small $H^0$}
\Text(20,30)[]{\small $H^0$}
\Text(-20,-27)[]{\small $H^0$}
\Text(25,-27)[]{\small $H^0$}
\end{picture}
\begin{picture}(100,70)(-50,-35)
\DashCArc(0,0)(15,0,360){3}
\DashLine(0,15)(-10,25){2}
\DashLine(0,15)(10,25){2}
\DashLine(-10.5,-10.5)(-20,-20){2}
\DashLine(10.5,-10.5)(20,-20){2}
\Text(-22,5)[]{\small $\str$}
\Text(22,5)[]{\small $\str$}
\Text(2,-23)[]{\small $\stl$}
\Text(-15,30)[]{\small $H^0$}
\Text(20,30)[]{\small $H^0$}
\Text(-20,-27)[]{\small $H^0$}
\Text(25,-27)[]{\small $H^0$}
\end{picture}
\begin{picture}(100,70)(-50,-35)
\DashCArc(0,0)(15,0,360){3}
\DashLine(0,15)(-10,25){2}
\DashLine(0,15)(10,25){2}
\DashLine(-10.5,-10.5)(-20,-20){2}
\DashLine(10.5,-10.5)(20,-20){2}
\Text(-22,5)[]{\small $\stl$}
\Text(22,5)[]{\small $\stl$}
\Text(2,-22)[]{\small $\scr$}
\Text(-15,30)[]{\small $H^0$}
\Text(20,30)[]{\small $H^0$}
\Text(-20,-27)[]{\small $H^0$}
\Text(25,-27)[]{\small $H^0$}
\end{picture}
\begin{picture}(100,70)(-50,-35)
\DashCArc(0,0)(15,0,360){3}
\DashLine(0,15)(-10,25){2}
\DashLine(0,15)(10,25){2}
\DashLine(-10.5,-10.5)(-20,-20){2}
\DashLine(10.5,-10.5)(20,-20){2}
\Text(-22,5)[]{\small $\str$}
\Text(22,5)[]{\small $\str$}
\Text(2,-22)[]{\small $\scl$}
\Text(-15,30)[]{\small $H^0$}
\Text(20,30)[]{\small $H^0$}
\Text(-20,-27)[]{\small $H^0$}
\Text(25,-27)[]{\small $H^0$}
\end{picture}
\end{center}
\begin{center}
\begin{picture}(88,70)(-44,-35)
\DashCArc(0,0)(15,0,360){3}
\DashLine(-10.5,10.5)(-20,20){2}
\DashLine(10.5,10.5)(20,20){2}
\DashLine(-10.5,-10.5)(-20,-20){2}
\DashLine(10.5,-10.5)(20,-20){2}
\Text(-22,3)[]{\small $\stl$}
\Text(22,3)[]{\small $\stl$}
\Text(2,22)[]{\small $\str$}
\Text(2,-23)[]{\small $\str$}
\Text(-20,27)[]{\small $H^0$}
\Text(25,27)[]{\small $H^0$}
\Text(-20,-27)[]{\small $H^0$}
\Text(25,-27)[]{\small $H^0$}
\end{picture}
\begin{picture}(88,70)(-44,-35)
\DashCArc(0,0)(15,0,360){3}
\DashLine(-10.5,10.5)(-20,20){2}
\DashLine(10.5,10.5)(20,20){2}
\DashLine(-10.5,-10.5)(-20,-20){2}
\DashLine(10.5,-10.5)(20,-20){2}
\Text(-22,3)[]{\small $\stl$}
\Text(22,3)[]{\small $\stl$}
\Text(2,21)[]{\small $\scr$}
\Text(2,-22)[]{\small $\scr$}
\Text(-20,27)[]{\small $H^0$}
\Text(25,27)[]{\small $H^0$}
\Text(-20,-27)[]{\small $H^0$}
\Text(25,-27)[]{\small $H^0$}
\end{picture}
\begin{picture}(88,70)(-44,-35)
\DashCArc(0,0)(15,0,360){3}
\DashLine(-10.5,10.5)(-20,20){2}
\DashLine(10.5,10.5)(20,20){2}
\DashLine(-10.5,-10.5)(-20,-20){2}
\DashLine(10.5,-10.5)(20,-20){2}
\Text(-22,3)[]{\small $\scl$}
\Text(22,3)[]{\small $\scl$}
\Text(2,22)[]{\small $\str$}
\Text(2,-23)[]{\small $\str$}
\Text(-20,27)[]{\small $H^0$}
\Text(25,27)[]{\small $H^0$}
\Text(-20,-27)[]{\small $H^0$}
\Text(25,-27)[]{\small $H^0$}
\end{picture}
\begin{picture}(88,70)(-44,-35)
\DashCArc(0,0)(15,0,360){3}
\DashLine(-10.5,10.5)(-20,20){2}
\DashLine(10.5,10.5)(20,20){2}
\DashLine(-10.5,-10.5)(-20,-20){2}
\DashLine(10.5,-10.5)(20,-20){2}
\Text(-22,3)[]{\small $\stl$}
\Text(22,3)[]{\small $\stl$}
\Text(2,22)[]{\small $\str$}
\Text(2,-22)[]{\small $\scr$}
\Text(-20,27)[]{\small $H^0$}
\Text(25,27)[]{\small $H^0$}
\Text(-20,-27)[]{\small $H^0$}
\Text(25,-27)[]{\small $H^0$}
\end{picture}
\begin{picture}(88,70)(-44,-35)
\DashCArc(0,0)(15,0,360){3}
\DashLine(-10.5,10.5)(-20,20){2}
\DashLine(10.5,10.5)(20,20){2}
\DashLine(-10.5,-10.5)(-20,-20){2}
\DashLine(10.5,-10.5)(20,-20){2}
\Text(-22,3)[]{\small $\stl$}
\Text(22,0)[]{\small $\scl$}
\Text(2,22)[]{\small $\str$}
\Text(2,-23)[]{\small $\str$}
\Text(-20,27)[]{\small $H^0$}
\Text(25,27)[]{\small $H^0$}
\Text(-20,-27)[]{\small $H^0$}
\Text(25,-27)[]{\small $H^0$}
\end{picture}
\end{center}
\caption{\em One-loop stop/scharm diagrams that contribute to the 
Higgs quartic coupling at ${\cal O}(y_t^4)$ through trilinear 
interactions.} 
\label{mhfig}
\end{figure}
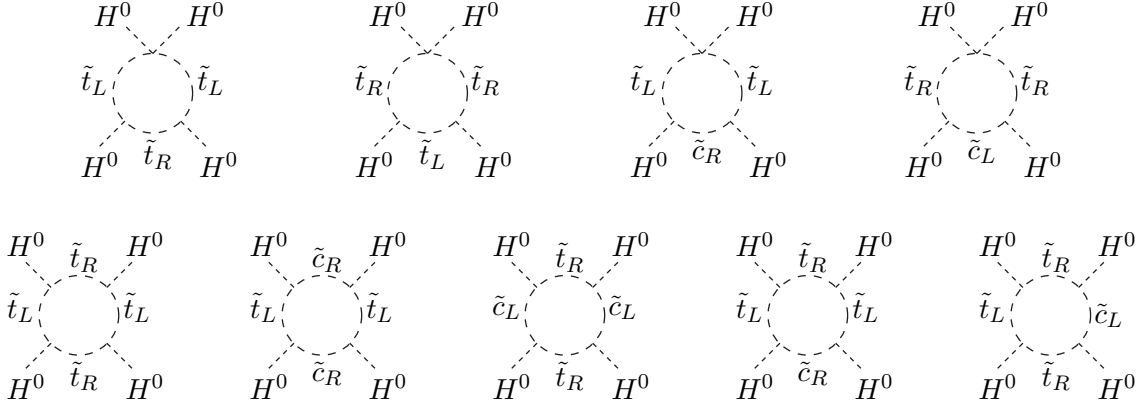

Our purpose in this Section is to generalize the one-loop
calculation of $\Delta$ by including the effect of the
flavour violating $A$-terms $\tca$ and $\cta$ of eq.~(\ref{vtril}), 
which couple the Higgs field to stop and scharm squarks. 
The one-loop stop/scharm diagrams that contribute to $\Delta$ 
at ${\cal O}(y_t^4)$ are shown in Fig.~\ref{mhfig}. 
%
Those in the first (second) row
are quadratic (quartic) in $X_t, \tca, \cta$
and give positive (negative) contributions to $\Delta$.  
We find: 
\bea
\label{delta}
\Delta & =  & 
\frac{\xxt}{\mmtl-\mmtr} \log \frac{\mmtl}{\mmtr}
+ \frac{\aact}{\mmcr} f_1 \! \left(\frac{\mmtl}{\mmcr}\right)
+ \frac{\aatc}{\mmcl} f_1 \! \left(\frac{\mmtr}{\mmcl}\right)
\nn \\
& & 
- \frac{1}{2}\left[ 
\frac{\xt4}{\mt^4_{t_L}} f_2 \! \left(\frac{\mmtr}{\mmtl}\right)
+ \frac{\act4}{\mt^4_{c_R}} f_2 \! \left(\frac{\mmtl}{\mmcr}\right)
+ \frac{\atc4}{\mt^4_{c_L}} f_2 \! \left(\frac{\mmtr}{\mmcl}\right)
\right]
\nn \\
& & 
-\frac{\xxt \aact}{\mmcr-\mmtr} 
\left[ \frac{1}{\mmtr} f_1\left(\frac{\mmtl}{\mmtr}\right)
-\frac{1}{\mmcr} f_1\left(\frac{\mmtl}{\mmcr}\right) \right]
\nn \\
& &
-\frac{\xxt \aatc}{\mmcl-\mmtl} 
\left[ \frac{1}{\mmtl} f_1\left(\frac{\mmtr}{\mmtl}\right)
-\frac{1}{\mmcl} f_1\left(\frac{\mmtr}{\mmcl}\right) \right]  \, ,
\eea
where $f_i(x)$ are positive functions 
[with $f_1(1)=\frac{1}{2}$, $f_2(1)=\frac{1}{6}$]:
\be
\label{f1f2}
f_1(x) =  -\frac{1}{(x-1)^2} \log x + \frac{1}{x-1}
\;\;\; , \;\;\; 
f_2(x) =  \frac{x+1}{(x-1)^3} \log x - \frac{2}{(x-1)^2} \, .
\ee
The overall size of the threshold function $\Delta$ as well as its sign
depend on the competition of positive and negative terms, in analogy
to the familiar case with $X_t$ only. The novel contributions 
induced by $\cta$ and $\tca$ can be of the same order
as the standard ones driven by $X_t$. As the total effect depends
on several mass parameters, a pre-fixed value of 
$\Delta$ is associated with some hypersurface in a 
multi-dimensional parameter space. To simplify the discussion, 
suppose that $(\stl, \str)$ have a common mass
$\mmt$ ($\simeq \mmtl \simeq \mmtr$) and that
$(\scl, \scr)$ have a common mass
$\mmc$ ($\simeq \mmcl \simeq \mmcr$). 
In this limit, the threshold function becomes:
\bea
\label{deltatc}
\Delta & = & 
\frac{\xxt}{\mmt} +
\frac{\aact+\aatc}{\mmc} f_1\left(\frac{\mmt}{\mmc}\right)
\nn \\
&- &
\left[ \frac{1}{12}\frac{\xt4}{\mt^4_t}
+ \frac{1}{2} \frac{\act4+\atc4}{\mt^4_c} f_2\left(\frac{\mmt}{\mmc}\right)
+ \frac{\xxt}{\mmt} \cdot
\frac{\aact+\aatc}{\mmc} f_3\left(\frac{\mmt}{\mmc}\right) \right] \; ,
\eea
where $f_3(x)$ is another positive function [with $f_3(1)=\frac{1}{6}$]:
\be
\label{f3}
f_3(x)  =  -\frac{x}{(x-1)^3} \log x + \frac{x+1}{2 (x-1)^2}  \, .
\ee
A further simplification occurs in the fully degenerate 
limit $\mmc = \mmt =\mt^2$:
\be
\label{deltam}
\Delta =\frac{\xxt}{\mt^2}+\frac{\aatc+\aact}{2 \, \mt^2}
-\frac{\xt4+\atc4+\act4 + 2 \xxt (\aatc+\aact)}
{12 \, \mt^4} \; .
\ee

Let us consider, for instance, the simplified expression in
eq.~(\ref{deltam}). Here $\Delta$ is a function of
only three dimensionless variables, namely 
$(x_t,a_{tc},a_{ct})\equiv (|X_t|/\mt,|\tca|/\mt,|\cta|/\mt)$. 
By a simple analytical study, we find the interesting result
that $\Delta$ is maximal at the `standard point' 
$(x_t,a_{tc},a_{ct})=(\sqrt{6},0,0)$, where $\Delta=3$. 
There are other extremal points where the flavour changing 
trilinears do not vanish, namely $(0,\sqrt{3},0)$, 
$(0,0,\sqrt{3})$ and $(0,\sqrt{3},\sqrt{3})$. At such extrema,
which are saddle points, $\Delta$ take values $3/4$, $3/4$ and
$3/2$, respectively. 
In a significant portion of the three-dimensional 
parameter space spanned by $(x_t,a_{tc},a_{ct})$
one can obtain $\Delta \simgt 1$. The role of $x_t$ is
crucial in order to reach $\Delta \simgt 2$.

Consider now a slightly more general scenario in which
stop and scharm masses are characterized by 
two distinct parameters $\mmt$ and $\mmc$, such that 
$\Delta$ is given by eq.~(\ref{deltatc}).
The parameter space can be described by three coordinates 
associated with the trilinear couplings,
which we take as $(x_t,a_{tc},a_{ct}) \equiv 
(|X_t|/\mt_t,|\tca|/\mt_c,|\cta|/\mt_c)$,
plus the ratio $r\equiv\mt_c/\mt_t$, which we
treat as an external parameter.
By an analytical study of $\Delta$ for fixed $r$,
we find the standard extremum at $(\sqrt{6},0,0)$
as well as other ones at $(0,a_*,0)$, $(0,0,a_*)$
and $(0,a_*,a_*)$, where $a_*=\sqrt{f_1/f_2}$ and 
$f_i \equiv f_i(1/r^2)$. Another extremum (a saddle point)
appears for $ 1 < r \simlt 5$. The extremum at $(0,a_*,a_*)$ 
is interesting because it is a local maximum for $r>1$. 
The associated value of $\Delta$ is  
$\Delta_*=f_1^2/f_2$, which increases for increasing $r$: 
for $r=(0.5;\, 1;\, 2;\, 3;\, 4;\, 5;\, 6)$ one finds  
$\Delta_* \simeq (0.9;\, 1.5;\, 2.3;\, 2.9;\, 3.4;\, 3.7;\, 4)$.
This behaviour follows from the mild (logarithmic) enhancement
of the coefficient functions $f_1$ and $f_2$, which is easily
interpreted through the diagrams in Fig.~\ref{mhfig}.
The other extrema  $(0,a_*,0)$ and $(0,0,a_*)$
are saddle points and have $\Delta=\frac{1}{2} \Delta_*$.
By comparing the reported values of $\Delta_*$ with $\Delta=3$ at 
the standard extremum $(\sqrt{6},0,0)$, we can see that
the latter point is no longer the absolute maximum for $r \simgt 3$.
For $r \simgt 5$, it is not even a local maximum and becomes
a saddle point (namely, $\Delta$ increases if ones moves away
from that point in the flavour violating directions).

\begin{figure}[tb]
\begin{center}
\vskip -3 cm
\psfrag{ r1 }{$\mt_c =  \mt_t$}
\psfrag{ r4 }{$\mt_c =  4\, \mt_t$}
\epsfig{file=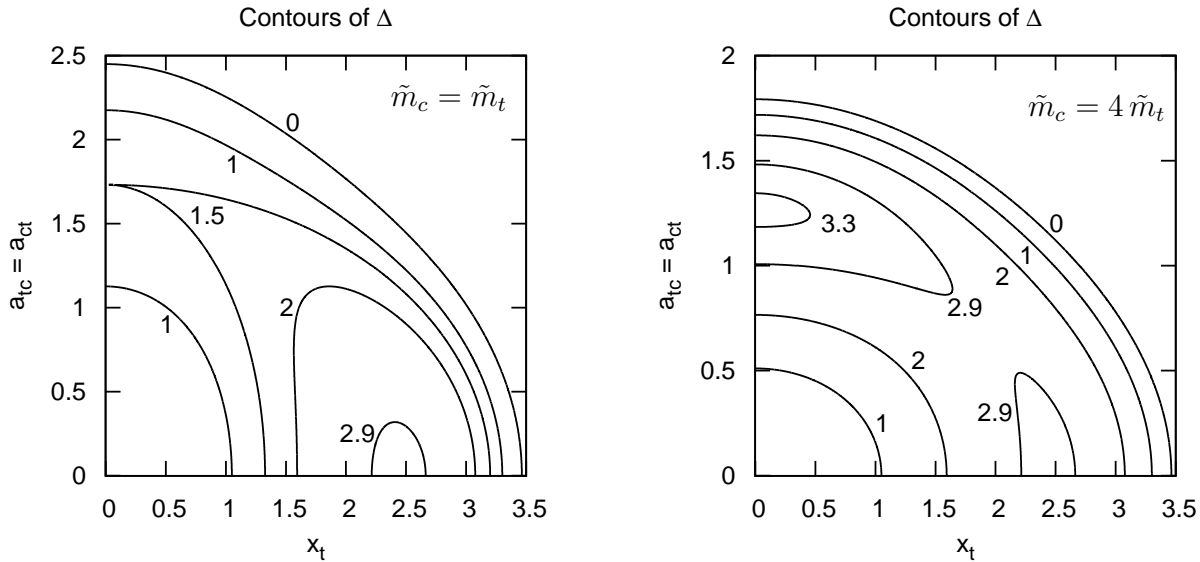,scale=0.68,angle=-90}
\vskip -2 cm
\caption{\em Iso-contours of $\Delta$ in the subspace spanned
by $x_t$ and $a_{tc}=a_{ct}$, for $r=1$ ($\mt_c=\mt_t$, left
panel) and $r=4$ ($\mt_c= 4\, \mt_t$, right panel).}
\label{deltafig}
\end{center}
\end{figure}

The behaviour of the threshold function $\Delta$ is further
illustrated in Fig.~\ref{deltafig}, where some iso-contours
are shown in a two-dimensional subspace spanned by $x_t$
and $a_{tc}=a_{ct}$, for either $r=1$ (left panel) or $r=4$ 
(right panel). The case $r=1$ (namely, $\mt_c=\mt_t$) is
the degenerate limit, already discussed above.
One can easily recognize the standard maximum
at $x_t=\sqrt{6}$ (and $a_{tc}=a_{ct}=0$), 
where $\Delta=3$, and the other extremum (saddle point) at 
$a_{tc}=a_{ct}=\sqrt{3}$ (and $x_t=0$), where $\Delta=1.5$.  
The case $r=4$  (namely, $\mt_c= 4\, \mt_t$) is an example
of a moderately hierarchical scenario. The extremum on the 
$a_{tc}=a_{ct}$ axis has turned into a maximum, and $\Delta$ is 
higher there than at the standard maximum on the $x_t$ axis. 
More generally, by comparing this case with the previous one,
one can notice the expansion of the region 
of parameter space where  $\Delta \simgt 2$, which translates 
into a linear correction to $m_h$ larger than about $10$~GeV.
As $\Delta\simgt 3$ can also be obtained in such hierarchical
scenarios, even shifts of ${\cal O}(20)$~GeV are possible.
In summary, significant positive threshold corrections to the Higgs
mass can be achieved in a variety of ways, by suitable 
combinations of the flavour conserving and flavour 
violating trilinear couplings. If any of such 
parameters is too large, though, the corrections quickly
become negative, since $\Delta$ is then dominated 
by negative quartic terms. In our examples in Fig.~\ref{deltafig}
this occurs to the right of the iso-contours where $\Delta=0$.
Such parameter regions are also disfavoured because
the tree-level potential can become unbounded from below along
coloured directions or develop colour-breaking minima 
\cite{ccb,casas}. 

A side remark may be added about the impact of sizeable values of
$\tca$ and/or $\cta$ on the naturalness of the weak scale. In fact,
although we have chosen to avoid discussing renormalization
effects above the SUSY scale, it should be mentioned that
the soft mass of the Higgs doublet $H_u$, {\it i.e.} $\mt^2_{H_u}$,  
receives logarithmic corrections proportional to 
$|A_t|^2+\aatc+\aact$. Therefore fine-tuning issues are not
alleviated by the presence of $\tca$ and $\cta$, 
particularly in case such parameters are of order $\mt_c$
and the latter is much larger than $\mt_t$. 

We conclude this Section by some comments on earlier results 
presented in the literature. The influence of flavour violating $A$-terms
on  $m_h$ was noticed in \cite{Cao:2006xb}
and confirmed in \cite{achhp}. However, such papers put a special
emphasis on the potentially large {\em negative} effect of
such trilinear couplings on the Higgs mass, which in fact was 
used to constrain their magnitude. On the other hand,
it was recently pointed out in \cite{Kowalska:2014opa} that a
sizeable {\em positive} effect on $m_h$ can also be achieved,
especially in the case of a hierarchical squark spectrum
($\mmc \gg \mmt$, in our notation). Our study confirms this
observation. Upon comparing our analytical results  
with those presented in \cite{Kowalska:2014opa}, though,
we have found agreement only in the degenerate case [$\mmc = \mmt$, 
eq.~(\ref{deltam})], not in the non-degenerate one  
[$\mmc \neq \mmt$, eq.~(\ref{deltatc})].


\section{The $\rho$ parameter}

In the previous Section we have examined certain SUSY corrections 
to the quartic operator $|H^0|^4$, which controls
the mass of the physical Higgs boson $h \subset H^0$. Other properties
of $h$ will be investigated in subsequent Sections. 
Here, instead, we will discuss the impact of SUSY
corrections to the $\rho$ parameter, where only the expectation
value $\langle |H^0| \rangle=v$ is relevant.
We select $\de \rho$ ($=\epsilon_1=\alpha \, \de T$) 
as the most representative quantity that affects 
electroweak precision observables, and recall that new
physics contributions to $\de \rho$ are constrained to be
at the per mille level at most \cite{gfitter}.
For instance, $\de \rho$ corrects the SM predictions
for the $W$ mass $m^2_W$ and the effective leptonic weak mixing angle 
$\sin^2 \theta^{\ell}_{\rm eff}$ by an amount
$\de m_W^2/m_W^2 \simeq - \de  \sin^2 \theta^{\ell}_{\rm eff} \, /s^2_w 
\simeq a \, \de\rho$, where $s^2_w \simeq 0.23$ and 
$a=c_w^2/(c_w^2-s_w^2)\simeq 1.4$. 
In particular, a positive $\de \rho \simeq 10^{-3}$ would induce 
$\de m_W \simeq 60\, {\rm MeV}$. This can be taken as a
maximal allowed shift, since the current deviation on $m_W$ is
$m_W^{\rm SM}-m_W^{\rm exp} \simeq - 25 \, {\rm MeV}$, with a one-sigma 
error of about $17 \, {\rm MeV}$. More stringent constraints on
$\de \rho$ are expected from future measurements at the LHC and at other 
proposed colliders. For general new physics, a full analysis 
should include other possible sources of corrections to
$m^2_W$ and $\sin^2 \theta^{\ell}_{\rm eff}$, such as the $S$ 
parameter. In our case, though, these effects are subleading
(they are at most ${\cal O}(g^2 y_t^2)$, therefore smaller 
than the ${\cal O}(y_t^4)$ effects associated with $\de \rho$).

\begin{figure}[b]
\vspace{0.3 cm}
\begin{center}
\begin{picture}(140,70)(-70,-35)
\DashCArc(0,0)(20,0,360){3}
\Photon(-20,-30)(-1,-20){1.5}{4}
\Photon(1,-20)(20,-30){1.5}{4}
\DashLine(10,17)(20,30){2}
\DashLine(-10,17)(-20,30){2}
\DashLine(3,19)(7,35){2}
\DashLine(-3,19)(-7,35){2}
\GOval(0,17)(3,15)(0){0.9}
\Text(-28,0)[]{\small $\sq_L^{\, i}$}
\Text(28,0)[]{\small $\sq_L^{\, i}$}
\Text(30,35)[]{\small $H^0$}
\Text(-25,35)[]{\small $H^0$}
\Text(12,44)[]{\small $H^0$}
\Text(-8,44)[]{\small $H^0$}
\Text(-32,-30)[]{\small $W^a$}
\Text(32,-30)[]{\small $W^a$}
\end{picture}
\begin{picture}(140,70)(-70,-35)
\DashCArc(0,0)(20,0,360){3}
\Photon(-34,-20)(-17,-10){1.5}{4}
\Photon(17,-10)(34,-20){1.5}{4}
\DashLine(10,17)(20,30){2}
\DashLine(-10,17)(-20,30){2}
\DashLine(3,19)(7,35){2}
\DashLine(-3,19)(-7,35){2}
\GOval(0,17)(3,15)(0){0.9}
\Text(-28,5)[]{\small $\sq_L^{\, i}$}
\Text(28,5)[]{\small $\sq_L^{\, i}$}
\Text(0,-30)[]{\small $\sq_L^{\prime \, i}$}
\Text(30,35)[]{\small $H^0$}
\Text(-25,35)[]{\small $H^0$}
\Text(12,44)[]{\small $H^0$}
\Text(-8,44)[]{\small $H^0$}
\Text(-48,-20)[]{\small $W^a$}
\Text(48,-20)[]{\small $W^a$}
\end{picture}
\begin{picture}(140,70)(-70,-35)
\DashCArc(0,0)(20,0,360){3}
\Photon(-35,0)(-20,0){1.5}{3}
\Photon(20,0)(35,0){1.5}{3}
\DashLine(3,18)(14,33){2}
\DashLine(-3,18)(-14,33){2}
\DashLine(3,-18)(14,-33){2}
\DashLine(-3,-18)(-14,-33){2}
\GOval(0,19)(3,12)(0){0.9}
\GOval(0,-19)(3,12)(0){0.9}
\Text(-25,13)[]{\small $\sq_L^{\, i}$}
\Text(25,13)[]{\small $\sq_L^{\, j}$}
\Text(-27,-13)[]{\small $\sq_L^{\prime \, i}$}
\Text(25,-13)[]{\small $\sq_L^{\prime \, j}$}
\Text(18,42)[]{\small $H^0$}
\Text(-16,42)[]{\small $H^0$}
\Text(18,-40)[]{\small $H^0$}
\Text(-16,-40)[]{\small $H^0$}
\Text(-48,0)[]{\small $W^a$}
\Text(48,0)[]{\small $W^a$}
\end{picture}
\end{center}
\vspace{-0.3cm}
\caption{\em Classes of one-loop squark diagrams that contribute to the
self-energies of $SU(2)$ vector bosons with four Higgs insertions. 
The ovals indicate that Higgs lines can be inserted in all possible
ways on squark propagators, through trilinear or quartic couplings.
Only diagrams of the third class contribute to $\delta \rho$.} 
\label{wwfig}
\end{figure}
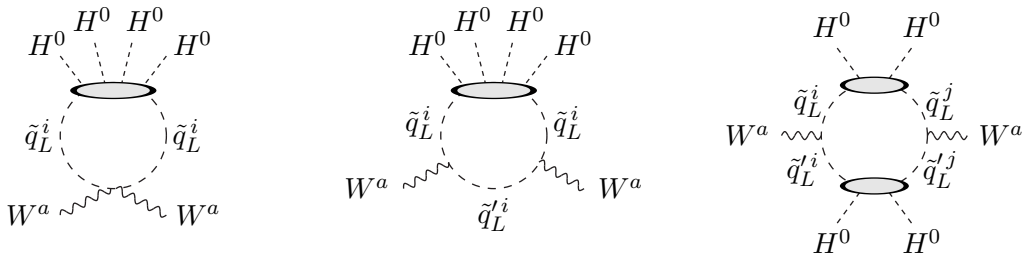

Squark contributions to $\de \rho$ can be evaluated through 
$\de \rho = [\Pi_{33}(0)-\Pi_{WW}(0)]/m_W^2$,
where $\Pi_{33}(0)$ and $\Pi_{WW}(0)$ are the 
self-energies of $W^3$ and $W^\pm$ at zero momentum
(up to the usual factor $g_{\mu\nu}$).
Diagrammatic computations have often been performed by diagonalizing
the squark mass matrices and keeping the full dependence
on $v$ \cite{rhobm,drho1,drho2}. As already declared, we choose 
to evaluate diagrams by explicitly inserting Higgs lines and 
looking for the leading non-vanishing terms in a $v^2$ expansion.
At ${\cal O}(v^0)$, both $\Pi_{33}(0)$ and $\Pi_{WW}(0)$
separately vanish by gauge invariance. At ${\cal O}(v^2)$,
those self-energies receive equal contributions, so there
is no net $\de \rho$ at this order. The leading  
contributions to $\de \rho$ arise from  ${\cal O}(v^4)$ terms 
in $W^a$ self-energies, so $\de \rho \sim v^2/\mt^2$. 
In fact, as well known, such corrections are associated with 
the $d=6$ effective operator $|H^{\dagger} D_{\mu} H|^2$:
if $c_{\rho}$ is the coefficient of the latter, 
$\de \rho \simeq - c_{\rho} v^2$.
As the relevant $W^a$ self-energies require (at least) four Higgs 
insertions, we find it very convenient to group the one-loop squark 
diagrams that contribute to them into three classes,
as shown schematically in Fig.~\ref{wwfig}.
%
%
The symbols $\sq_L^{\, i}, \sq_L^{\prime \, i}$ generically denote
up-type or down-type squarks of the $i$-th left-handed doublet, so 
it is understood that $\sq_L^{\, i}$ and $\sq_L^{\prime \, i}$ are
equal (different) if they couple to $W^3$ ($W^{\pm}$).
It is both useful and not restrictive to 
consider a basis where the $3\times 3$ mass matrix $\mt_Q^2$ of $SU(2)$ 
doublets is diagonal, so flavour transitions can only occur at 
Higgs vertices or in $\sq_R$ propagators. 
Consider the first class of diagrams in Fig.~\ref{wwfig}.
Each bilinear $|\sq_L^{\, i}|^2$ has the same quartic coupling to 
$|W^{\pm}|^2$ and to $\frac{1}{2}(W^3)^2$, by $SU(2)$ invariance.
This implies that there are equal contributions to $\Pi_{33}(0)$ 
and $\Pi_{WW}(0)$,
so the net contribution to $\de \rho$ is zero. A similar argument
can be applied to the second class of diagrams. Again, the
$SU(2)$ properties of the $W^a$-$\sq_L^{\, i}$-$\sq_L^{\prime \, i}$ 
vertices imply equal contributions to $\Pi_{33}(0)$ and $\Pi_{WW}(0)$
from each doublet, hence zero contribution to $\de \rho$. Thus we 
conclude that {\em only diagrams in the third class are relevant 
to} $\de \rho$. 

\begin{figure}[tb]
\begin{center}
\begin{picture}(200,70)(-70,-35)
\DashCArc(0,0)(15,0,360){3}
\Photon(-30,0)(-15,0){1.5}{3}
\Photon(15,0)(30,0){1.5}{3}
\DashLine(0,15)(-10,25){2}
\DashLine(0,15)(10,25){2}
\DashLine(0,-15)(-10,-25){2}
\DashLine(0,-15)(10,-25){2}
\Text(-7,2)[]{\small $\stl$}
\Text(9,2)[]{\small $\stl$}
\Text(-15,30)[]{\small $H^0$}
\Text(20,30)[]{\small $H^0$}
\Text(-15,-32)[]{\small $H^0$}
\Text(15,-32)[]{\small $H^0$}
\Text(-30,8)[]{\small $W^3$}
\Text(35,8)[]{\small $W^3$}
\end{picture}
\begin{picture}(88,70)(-20,-35)
\DashCArc(0,0)(15,0,360){3}
\Photon(-30,0)(-15,0){1.5}{3}
\Photon(15,0)(30,0){1.5}{3}
\DashLine(0,15)(-10,25){2}
\DashLine(0,15)(10,25){2}
\DashLine(-10.5,-10.5)(-20,-20){2}
\DashLine(10.5,-10.5)(20,-20){2}
\Text(-7,2)[]{\small $\stl$}
\Text(9,2)[]{\small $\stl$}
\Text(2,-23)[]{\small $\str$}
\Text(-15,30)[]{\small $H^0$}
\Text(20,30)[]{\small $H^0$}
\Text(-20,-27)[]{\small $H^0$}
\Text(25,-27)[]{\small $H^0$}
\Text(-30,8)[]{\small $W^3$}
\Text(35,8)[]{\small $W^3$}
\end{picture}
\begin{picture}(88,70)(-20,-35)
\DashCArc(0,0)(15,0,360){3}
\Photon(-30,0)(-15,0){1.5}{3}
\Photon(15,0)(30,0){1.5}{3}
\DashLine(0,15)(-10,25){2}
\DashLine(0,15)(10,25){2}
\DashLine(-10.5,-10.5)(-20,-20){2}
\DashLine(10.5,-10.5)(20,-20){2}
\Text(-7,2)[]{\small $\stl$}
\Text(9,2)[]{\small $\stl$}
\Text(2,-22)[]{\small $\scr$}
\Text(-15,30)[]{\small $H^0$}
\Text(20,30)[]{\small $H^0$}
\Text(-20,-27)[]{\small $H^0$}
\Text(25,-27)[]{\small $H^0$}
\Text(-30,8)[]{\small $W^3$}
\Text(35,8)[]{\small $W^3$}
\end{picture}
\end{center}
\begin{center}
\begin{picture}(88,70)(-44,-35)
\DashCArc(0,0)(15,0,360){3}
\Photon(-30,0)(-15,0){1.5}{3}
\Photon(15,0)(30,0){1.5}{3}
\DashLine(-10.5,10.5)(-20,20){2}
\DashLine(10.5,10.5)(20,20){2}
\DashLine(-10.5,-10.5)(-20,-20){2}
\DashLine(10.5,-10.5)(20,-20){2}
\Text(-7,2)[]{\small $\stl$}
\Text(9,2)[]{\small $\stl$}
\Text(2,23)[]{\small $\str$}
\Text(2,-23)[]{\small $\str$}
\Text(-20,27)[]{\small $H^0$}
\Text(25,27)[]{\small $H^0$}
\Text(-20,-27)[]{\small $H^0$}
\Text(25,-27)[]{\small $H^0$}
\Text(-30,8)[]{\small $W^3$}
\Text(35,8)[]{\small $W^3$}
\end{picture}
\begin{picture}(88,70)(-44,-35)
\DashCArc(0,0)(15,0,360){3}
\Photon(-30,0)(-15,0){1.5}{3}
\Photon(15,0)(30,0){1.5}{3}
\DashLine(-10.5,10.5)(-20,20){2}
\DashLine(10.5,10.5)(20,20){2}
\DashLine(-10.5,-10.5)(-20,-20){2}
\DashLine(10.5,-10.5)(20,-20){2}
\Text(-7,2)[]{\small $\stl$}
\Text(9,2)[]{\small $\stl$}
\Text(2,23)[]{\small $\scr$}
\Text(2,-22)[]{\small $\scr$}
\Text(-20,27)[]{\small $H^0$}
\Text(25,27)[]{\small $H^0$}
\Text(-20,-27)[]{\small $H^0$}
\Text(25,-27)[]{\small $H^0$}
\Text(-30,8)[]{\small $W^3$}
\Text(35,8)[]{\small $W^3$}
\end{picture}
\begin{picture}(88,70)(-44,-35)
\DashCArc(0,0)(15,0,360){3}
\Photon(-30,0)(-15,0){1.5}{3}
\Photon(15,0)(30,0){1.5}{3}
\DashLine(-10.5,10.5)(-20,20){2}
\DashLine(10.5,10.5)(20,20){2}
\DashLine(-10.5,-10.5)(-20,-20){2}
\DashLine(10.5,-10.5)(20,-20){2}
\Text(-7,2)[]{\small $\scl$}
\Text(8,1)[]{\small $\scl$}
\Text(2,23)[]{\small $\str$}
\Text(2,-23)[]{\small $\str$}
\Text(-20,27)[]{\small $H^0$}
\Text(25,27)[]{\small $H^0$}
\Text(-20,-27)[]{\small $H^0$}
\Text(25,-27)[]{\small $H^0$}
\Text(-30,8)[]{\small $W^3$}
\Text(35,8)[]{\small $W^3$}
\end{picture}
\begin{picture}(88,70)(-44,-35)
\DashCArc(0,0)(15,0,360){3}
\Photon(-30,0)(-15,0){1.5}{3}
\Photon(15,0)(30,0){1.5}{3}
\DashLine(-10.5,10.5)(-20,20){2}
\DashLine(10.5,10.5)(20,20){2}
\DashLine(-10.5,-10.5)(-20,-20){2}
\DashLine(10.5,-10.5)(20,-20){2}
\Text(-7,2)[]{\small $\stl$}
\Text(9,2)[]{\small $\stl$}
\Text(2,23)[]{\small $\str$}
\Text(2,-22)[]{\small $\scr$}
\Text(-20,27)[]{\small $H^0$}
\Text(25,27)[]{\small $H^0$}
\Text(-20,-27)[]{\small $H^0$}
\Text(25,-27)[]{\small $H^0$}
\Text(-30,8)[]{\small $W^3$}
\Text(35,8)[]{\small $W^3$}
\end{picture}
\begin{picture}(88,70)(-44,-35)
\DashCArc(0,0)(15,0,360){3}
\Photon(-30,0)(-15,0){1.5}{3}
\Photon(15,0)(30,0){1.5}{3}
\DashLine(-10.5,10.5)(-20,20){2}
\DashLine(10.5,10.5)(20,20){2}
\DashLine(-10.5,-10.5)(-20,-20){2}
\DashLine(10.5,-10.5)(20,-20){2}
\Text(-7,2)[]{\small $\stl$}
\Text(8,1)[]{\small $\scl$}
\Text(2,23)[]{\small $\str$}
\Text(2,-23)[]{\small $\str$}
\Text(-20,27)[]{\small $H^0$}
\Text(25,27)[]{\small $H^0$}
\Text(-20,-27)[]{\small $H^0$}
\Text(25,-27)[]{\small $H^0$}
\Text(-30,8)[]{\small $W^3$}
\Text(35,8)[]{\small $W^3$}
\end{picture}
\end{center}
\vspace{-0.5cm}
\caption{\em One-loop stop/scharm diagrams that contribute to the
$\rho$ parameter at ${\cal O}(y_t^4)$.} 
\label{rhofig}
\end{figure}
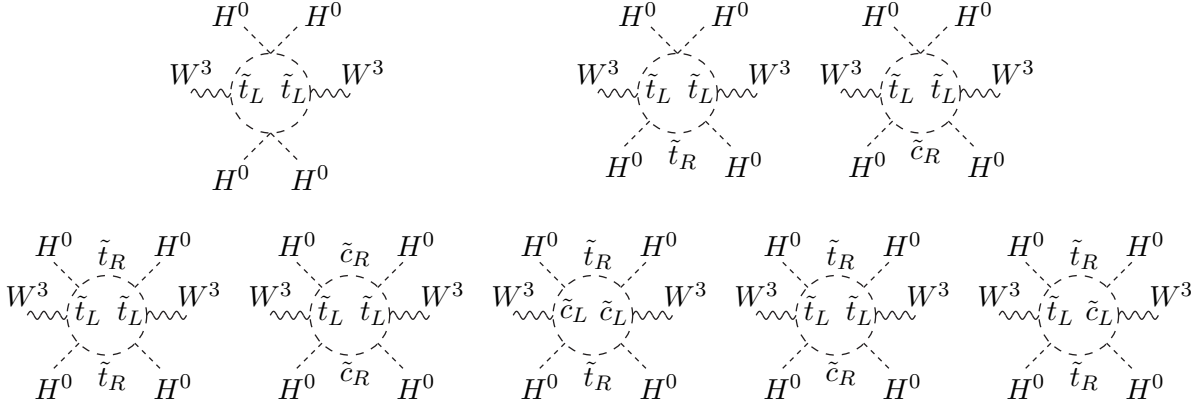

Our previous classification and conclusion 
are very general and go beyond the specific framework we are
interested in. Let us now specialize all that to our case.
Since we are looking for leading effects in $y_t$, the Higgs insertions
in the third class of diagrams in Fig.~\ref{wwfig} should involve 
$\stl$ or $\scl$, so the external gauge bosons are $W^3$-$W^3$. 
The stop/scharm diagrams that contribute to the
$\rho$ parameter at ${\cal O}(y_t^4)$ are shown
in Fig.~\ref{rhofig}.
%
%
Apart from the first diagram, the other ones 
are related to most of those in Fig.~\ref{mhfig}, namely, those
where two external $W^3$ bosons can be attached to two 
{\em distinct} $\sq_L$ propagators. 
The first diagram in Fig.~\ref{rhofig} gives a positive contribution
to $\de \rho$. The other ones in the first (second) row
are quadratic (quartic) in trilinear couplings and give negative 
(positive) contributions. The result of our computation is:
\bea
\label{deltarho}
\delta \rho & = &  
\frac{3 y_t^4}{16 \pi^2} \, \frac{ v^2 }{\mmtl} \left\{
\frac{1}{6} 
- \frac{\xxt}{\mmtr} g_1 \! \left(\frac{\mmtl}{\mmtr}\right)
- \frac{\aact}{\mmcr} g_1 \! \left(\frac{\mmtl}{\mmcr}\right)
\right.
\nn \\
& &
+ \frac{\xt4}{\mt^4_{t_R}} g_2 \! \left(\frac{\mmtl}{\mmtr}\right)
+ \frac{\act4}{\mt^4_{c_R}} g_2 \! \left(\frac{\mmtl}{\mmcr}\right)
+ \frac{\atc4}{\mt^4_{c_L}} \cdot \frac{\mmtl}{\mmcl}
 g_2 \! \left(\frac{\mmtr}{\mmcl}\right)
\nn \\
& &
+ \frac{\xxt \aact}{\mmcr-\mmtr} \left[
\frac{1}{\mmtr} g_1 \! \left(\frac{\mmtl}{\mmtr}\right)
-\frac{1}{\mmcr} g_1 \! \left(\frac{\mmtl}{\mmcr}\right)\right] 
\nn \\
& &
+2 \frac{\xxt \aatc}{(\mmcl-\mmtl)^2} \left[
f_3 \! \left(\frac{\mmtr}{\mmtl}\right)
+\frac{\mmtl}{\mmcl} f_3 \! \left(\frac{\mmtr}{\mmcl}\right)
\right.
\nn \\
& &
\left. \left.
+\frac{\mmtl}{\mmcl-\mmtl}\left(
 f_1 \! \left(\frac{\mmtr}{\mmtl}\right)
- f_1 \! \left(\frac{\mmtr}{\mmcl}\right) \right) \right]
\right\} \; ,
\eea
where $f_1(x)$ and $f_3(x)$ have been defined in eqs.~(\ref{f1f2}) 
and (\ref{f3}), and $g_i(x)$ are other positive functions [with 
$g_1(0)=\frac{1}{3}$, $g_1(1)=\frac{1}{12}$, $g_2(0)=\frac{1}{6}$, 
$g_2(1)=\frac{1}{60}$]:
\be
\label{g1g2}
g_1(x)  =  \frac{x}{(x-1)^4} \log x + 
\frac{x^2 - 5x - 2}{6 (x-1)^3}
\;\;\;  , \;\;\; 
g_2(x)  =  -\frac{x(x+1)}{(x-1)^5} \log x +
\frac{x^2 + 10 x +1}{6 (x-1)^4} \; .
\ee
The structure of eq.~(\ref{deltarho}) resembles that of $\Delta$
in eq.~(\ref{delta}). Flavour conserving and flavour changing trilinears 
appear on the same footing and can give effects of the 
same order. At variance with $\Delta$, though, $\de \rho$ is
not left-right symmetric and is suppressed by an overall factor 
$v^2/\mmtl$. All such features are expected, of course.
In the simplified scenario where $\mmtl \simeq \mmtr \simeq \mmt$
and $\mmcl \simeq \mmcr \simeq \mmc$, the above result reads:
\bea
\label{rhotc}
\delta \rho & = &  
\frac{3 y_t^4}{16 \pi^2} \, \frac{ v^2 }{\mmt}
\left\{
\frac{1}{6} 
- \frac{1}{12}\frac{\xxt}{\mmt}
- \frac{\aact}{\mmc} g_1 \! \left(\frac{\mmt}{\mmc}\right)
\right.
\nn \\
& &
+ \frac{1}{60}\frac{\xt4}{\mt^4_t} 
+ \left( \frac{\act4}{\mt^4_c} 
+ \frac{\atc4}{\mt^4_c} \cdot \frac{\mmt}{\mmc} \right)
 g_2 \! \left(\frac{\mmt}{\mmc}\right)
\nn \\
& & \left.
+ \frac{\xxt}{\mmt} \left[ \frac{\aact}{\mmc}
g_3 \! \left(\frac{\mmt}{\mmc}\right)
+ 2 \frac{\aatc}{\mmc} \cdot \frac{\mmt}{\mmc}  
\, g_2 \! \left(\frac{\mmt}{\mmc}\right) \right]
\right\} \; ,
\eea
where $g_3(x)$ is another positive function 
[with $g_3(0)=\frac{1}{12}$, $g_3(1)=\frac{1}{30}$]:
\be
g_3(x) =  \frac{x^2}{(x-1)^5} \log x +
\frac{x^3 -7 x^2 - 7 x +1}{12 (x-1)^4} \; .
\ee

In the flavour conserving limit ($\tca=\cta=0$), eq.~(\ref{rhotc}) reduces to
\be
\label{rhofc}
\delta \rho |_{\rm fl. cons.}  =   
\frac{y_t^4}{32 \pi^2} \, \frac{ v^2 }{\mmt}
\left[ 1 - \frac{1}{2}\frac{\xxt}{\mmt}
+ \frac{1}{10}\frac{\xt4}{\mt^4_t}  \right] \; ,
\ee
which is consistent with the first SUSY computation of $\de \rho$ \cite{rhobm}.
The expression of $\delta \rho |_{\rm fl. cons.}$ for $\mmtl \neq \mmtr$ 
can be easily read off from eq.~(\ref{deltarho}).
Eq.~(\ref{rhofc}) also agrees with one of the results presented in 
ref.~\cite{hlm}, where the coefficients of several $d=6$ effective operators 
were computed, in the flavour conserving case with degenerate stop 
masses\footnote{In ref.~\cite{hlm} subleading terms 
of order $g^2 y_t^2$ and $g^4$ have been evaluated as well. 
We confirm those terms, which we have obtained by including 
$D$-term contributions to both $\Pi_{33}(0)$ and $\Pi_{WW}(0)$, again
by considering diagrams of the third class in Fig.~\ref{wwfig},
with up-type and down-type squark propagators.  
In fact, we have found that even our general result for $\de \rho$ in 
eq.~(\ref{deltarho}) can easily be extended to account for all $D$-term 
effects, also including squarks and sleptons of all generations. 
The recipe is: {\em i)} in the term without trilinears,
replace $3 y_t^4/\mmtl$ by $3 ( y_t^2 +\cos 2\beta \, g^2/2 )^2/\mmtl+
(\cos 2\beta \, g^2/2 )^2 (3/\mmcl+3/\mmul+1/\mmel+1/\mmmul+1/\mmtaul)$;
{\em ii)} in the terms proportional to $\xxt$ and $\aact$,
replace $y_t^4$ by $y_t^2 ( y_t^2 +\cos 2\beta \, g^2/2 )$;
{\em iii)} add the term $-v^2 /(16 \pi^2) \cdot 
(3 y_t^2 \cos 2\beta \, g^2/2 )\aatc/(\mmcl\mmtr) \cdot g_1(\mmcl/\mmtr)$.}.

Other limits of eq.~(\ref{rhotc}) lead to simple expressions for $\de \rho$.
For instance, in the degenerate limit ($\mmc = \mmt =\mt^2$) 
we obtain:
\be
\label{rhodeg}
\delta \rho  =  \frac{y_t^4}{32 \pi^2} \, \frac{v^2}{\mt^2}
\left[ 1 - \frac{\xxt +\aact}{2\, \mt^2} 
+ \frac{\xt4+\atc4+\act4
+ 2 \xxt (\aatc+\aact)}{10 \, \mt^4} \right] \; .
\ee
The expression in brackets, which resembles $\Delta$ in 
eq.~(\ref{deltam}), is positive and generically ${\cal O}(1)$. 
More precisely, its value is $1$ at vanishing trilinears, 
minimal ($=\frac{3}{8}$) at $|X_t|^2+|\cta|^2=\frac{5}{2}\mt^2$ 
(with $|\tca|=0$), and large ($\gg 1$) if any of the trilinears
is much larger than $\mt$. Barring the latter case,
we can see that $\delta \rho$ is sufficiently suppressed
even for light squark masses (the prefactor 
$y_t^4 v^2/(32 \pi^2 \mt^2)$ is about $2 \cdot 10^{-4}$
for $\mt \sim 500$ GeV, if one takes $y_t^4 \sim 0.6$ 
at that scale). On the other hand, either eq.~(\ref{rhodeg})
or the more general expressions presented above,
eqs.~(\ref{deltarho}) and (\ref{rhotc}), could be
useful in case future electroweak precision measurements 
should require a small non-vanishing $\de \rho$.

Another interesting expression can be obtained from 
eq.~(\ref{rhotc}) in the hierarchical limit ($\mmc \gg \mmt$):
\be
\label{rhohier}
\delta \rho \simeq \frac{y_t^4}{32 \pi^2} \, \frac{v^2}{\mmt}
\left[\left( 1 -\frac{\aact}{\mmc}\right)^{\! 2} - 
\frac{1}{2} \frac{\xxt}{\mmt} \left( 1 -\frac{\aact}{\mmc}\right)
+ \frac{1}{10} \frac{\xt4}{\mt^4_t}  \right] \; .
\ee
Here flavour violating parameters only appear through $\aact/\mmc$, 
which should be actually interpreted as $\aact/\mmcr$. 
Terms dependent on $\aatc$ and $\mmcl$ are suppressed
by $\mmt/\mmcl$ and are not shown. The previous expression
is consistent with the result one obtains by first decoupling $\scr$
and then computing $\de \rho$. In this approach, the 
trilinear couplings $y_t\cta H^0 \scr^{\,*} \stl+{\rm h.c.}$ 
and the tree-level exchange of $\scr$ generate 
an effective quartic interaction of the same form
$|H^0|^2 |\stl|^2$ as the SUSY one, such that  
the overall effective coupling is $ \xi y_t^2  |H^0|^2 |\stl|^2$, 
where $\xi\equiv 1 - \aact/\mmcr$. Therefore one can take 
the expression of $\de \rho$ in the flavour conserving case, 
eq.~(\ref{rhofc}), rescale the first term by $\xi^2$ and the second one
by $\xi$, since they originate  
from diagrams with either two or one insertion(s)
of $|H^0|^2 |\stl|^2$, respectively. In this way 
eq.~(\ref{rhohier}) is recovered. As far as the size of $\de \rho$
is concerned, we can notice again that eq.~(\ref{rhohier}) exhibits a
suppression factor, controlled by $v^2/\mmt$,
times an ${\cal O}(1)$ factor, {\em i.e.},
the expression in square brackets. The latter one is positive
except at $\aact=\mmc$ and $X_t=0$, where it
vanishes, so $\de \rho$ is further suppressed in 
a neighbourhood of that point. Notice that scharm effects
are significant even for $\mmc \gg \mmt$, provided
$\aact= {\cal O}(\mmc)$.


\section{The processes $g g \leftrightarrow h$ and $h \to \ga \ga$}

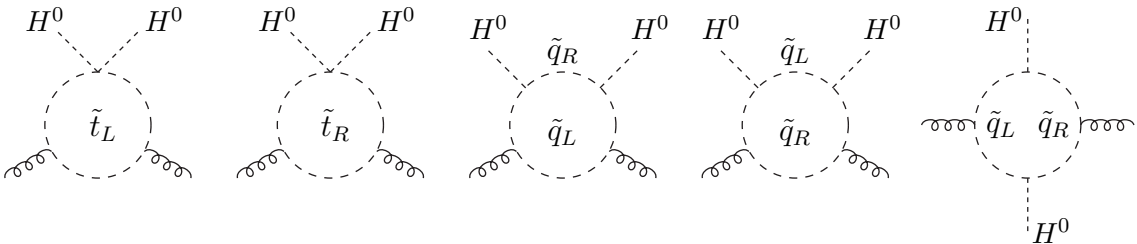
\begin{figure}[b]
\vspace{0.5cm}
\begin{center}
\begin{picture}(84,70)(-42,-30)
\DashCArc(0,0)(20,0,360){3}
\Gluon(-35,-20)(-17.5,-10){2}{3}
\Gluon(17.5,-10)(35,-20){2}{3}
\DashLine(0,20)(-15,35){2}
\DashLine(0,20)(15,35){2}
\Text(0,0)[]{ $\stl$}
\Text(-20,40)[]{\small $H^0$}
\Text(25,40)[]{\small $H^0$}
\end{picture}
\begin{picture}(84,70)(-42,-30)
\DashCArc(0,0)(20,0,360){3}
\Gluon(-35,-20)(-17.5,-10){2}{3}
\Gluon(17.5,-10)(35,-20){2}{3}
\DashLine(0,20)(-15,35){2}
\DashLine(0,20)(15,35){2}
\Text(0,0)[]{ $\str$}
\Text(-20,40)[]{\small $H^0$}
\Text(25,40)[]{\small $H^0$}
\end{picture}
\begin{picture}(84,70)(-42,-30)
\DashCArc(0,0)(20,0,360){3}
\Gluon(-35,-20)(-17.5,-10){2}{3}
\Gluon(17.5,-10)(35,-20){2}{3}
\DashLine(-14,14)(-28,28){2}
\DashLine(14,14)(28,28){2}
\Text(0,-3)[]{$\sq_L$}
\Text(0,28)[]{$\sq_R$}
\Text(-28,37)[]{\small $H^0$}
\Text(33,37)[]{\small $H^0$}
\end{picture}
\begin{picture}(84,70)(-42,-30)
\DashCArc(0,0)(20,0,360){3}
\Gluon(-35,-20)(-17.5,-10){2}{3}
\Gluon(17.5,-10)(35,-20){2}{3}
\DashLine(-14,14)(-28,28){2}
\DashLine(14,14)(28,28){2}
\Text(0,-3)[]{$\sq_R$}
\Text(0,28)[]{$\sq_L$}
\Text(-28,37)[]{\small $H^0$}
\Text(33,37)[]{\small $H^0$}
\end{picture}
\begin{picture}(84,70)(-42,-30)
\DashCArc(0,0)(20,0,360){3}
\Gluon(-40,0)(-20,0){2}{3}
\Gluon(20,0)(40,0){2}{3}
\DashLine(0,20)(0,40){2}
\DashLine(0,-20)(0,-40){2}
\Text(-10,0)[]{$\sq_L$}
\Text(10,0)[]{$\sq_R$}
\Text(-9,40)[]{\small $H^0$}
\Text(9,-40)[]{\small $H^0$}
\end{picture}
\end{center}
\caption{\em One-loop stop/scharm diagrams that contribute to the
effective Higgs coupling to gluons or photons at 
${\cal O}(y_t^2)$ ($\sq_L \in \{ \stl,\scl \}$, 
$\sq_R \in \{ \str,\scr \}$).} 
\label{ggfig}
\end{figure}

After the previous digression on the $\rho$ parameter, we now return
to discuss properties of the physical Higgs boson $h$, namely
its effective couplings with other SM particles. 
Such couplings can be parametrized through phenomenological scale 
factors $\kappa_i$ \cite{LHCwg}, which encode possible
deviations from the SM predictions. In particular, $\kappa_g$ 
and $\kappa_\ga$ are associated with crucial processes such as 
$gg \leftrightarrow h$ and $h\to \ga \ga$:
\be 
\frac{\sigma(gg \to h)}{\sigma^{\rm SM}(gg\to h)} \simeq
\frac{\Gamma(h\to gg)}{\Gamma^{\rm SM}(h\to gg)}=
\kappa_g^2 
\;\; , \;\;\;\;\;\;\;
\frac{\Gamma(h\to \ga \ga)}{\Gamma^{\rm SM}(h\to \ga \ga)} =
\kappa_{\ga}^2  \; .
\ee
We can write $\kappa_g=1+\de \kappa_g$ and 
$\kappa_{\ga}=1+\de \kappa_{\ga}$, where  
$\de \kappa_g$ and $\de \kappa_{\ga}$ encode the corrections
from new physics, normalized to the SM amplitudes
($\de \kappa_g=\delta {\cal A}_{hgg}/{\cal A}^{\rm SM}_{hgg} , \,
\de \kappa_\ga=\delta {\cal A}_{h\ga\ga}/{\cal A}^{\rm SM}_{h\ga\ga}$).
In our scenario and within our assumptions, 
$\de \kappa_g$ and $\de \kappa_{\ga}$ are proportional to $v^2/\mt^2$
and are related to the $d=6$ operators $|H^0|^2 G_{\mu\nu} G^{\mu\nu}$
and $|H^0|^2 F_{\mu\nu} F^{\mu\nu}$, which receive ${\cal O}(y_t^2)$
contributions from the one-loop stop/scharm diagrams shown in 
Fig.~\ref{ggfig}.
%
%
Let us consider first the case of external gluons.
We have computed those diagrams at vanishing Higgs momenta and kept
terms quadratic in the gluon momenta (the terms at zero gluon momenta 
are cancelled by other diagrams with quartic gluon-squark
couplings, consistently with gauge invariance).
By comparing with the SM result, which is dominated by a top
loop, we find:
\be
\label{kg}
\de \kappa_g \simeq
\frac{m_t^2}{4}\left[ 
\frac{1}{\mmtl}\left(1-\frac{\aact}{\mmcr}\right)
+\frac{1}{\mmtr}\left(1-\frac{\aatc}{\mmcl}\right)
-\frac{\xxt}{\mmtl \mmtr} \right]  \; .
\ee
We have also checked that the same expression can be derived through 
Higgs low-energy theorems \cite{hlt}. Indeed, the one-loop correction 
to the coefficient of 
$G_{\mu\nu} G^{\mu\nu}$ induced by stop and scharm squarks in a Higgs 
background is proportional to $b \log \det {\cal M}^2$, 
where $b$ is the appropriate $\beta$-function coefficient and 
${\cal M}^2={\cal M}^2(H^0)$ is the ($4\times4$)
Higgs-dependent squark mass matrix. Therefore, by expanding 
$\log \det {\cal M}^2$ up to ${\cal O}(|H^0|^2)$,
we have found the coupling of interest and recovered 
eq.~(\ref{kg}) after normalizing to the top contribution. 
Our result for $\de \kappa_g$ generalizes
the well studied one without flavour violation (see, {\em e.g.},
\cite{dj98,djou,stops,hlm}).
The effect of the novel terms proportional to $\aact$ and $\aatc$ 
is analogous to that of $\xxt$, {\em i.e.}, all trilinear
parameters generate negative contributions to $\de \kappa_g$, 
which can therefore have either sign.
Although eq.~(\ref{kg}) is already very simple, for completeness
we also write  $\de \kappa_g$ in the simplified scenario
where $\mmtl \simeq \mmtr \simeq \mmt$
and $\mmcl \simeq \mmcr \simeq \mmc$:
\be
\label{kg1}
\de \kappa_g \simeq \frac{m_t^2}{2 \, \mmt}\left[1 - \frac{1}{2}\left(
\frac{\xxt}{\mmt} +\frac{\aact + \aatc}{\mmc}\right) \right] \; .
\ee
Notice that for $\mmc \simgt \mmt$ 
the size of $\de \kappa_g$ is controlled by $m_t^2/\mmt$, 
yet the effects of scharm squarks are not sub-leading even 
in the hierarchical limit ($\mmc \gg \mmt$), 
provided $\aatc$ and/or $\aact$ are of order $\mmc$.  
An analogous comment applies to the general result of eq.~(\ref{kg}).
Finally, in the degenerate limit ($\mmc = \mmt =\mt^2$) the
previous result becomes:
\be
\label{kg2}
\de \kappa_g \simeq \frac{m_t^2}{2 \, \mt^2}\left[1 - 
\frac{\xxt +\aact + \aatc}{2 \, \mt^2}  \right]  \; .
\ee

The computation of the squark contribution to the Higgs-photon coupling  
is completely analogous to that of the Higgs-gluon coupling. The main
change is the normalization to the SM amplitude, where the leading
one-loop effect comes from $W$'s whilst the top loop
generates a smaller contribution of opposite sign. In practice, since 
${\cal A}^{\rm top}_{h\ga\ga}\simeq -0.3 \, {\cal A}^{\rm SM}_{h\ga\ga}$,
one gets $\de \kappa_{\ga} \simeq -0.3 \, \de \kappa_g$.

The latter (anti)correlation between $\de \kappa_{\ga}$ and $\de \kappa_g$
holds when other SUSY contributions to $\de \kappa_{\ga}$ are
negligible. That relation 
is useful also because LHC results on Higgs physics are sometimes 
presented as confidence regions in the plane $(\kappa_{\ga},\kappa_g)$,
under the assumption that other Higgs couplings are SM-like.
Therefore we can intersect the line 
$\de \kappa_{\ga} = -0.3 \, \de \kappa_g$
with the $95\% $ C.L. contours reported by either CMS \cite{cms14} 
or ATLAS \cite{atlas15} and infer bounds such as  
$-0.3 \simlt \de \kappa_g \simlt 0.15$ or
$-0.15 \simlt \de \kappa_g \simlt 0.45$, respectively.
These ranges translate into constraints on the combination of 
masses and trilinear couplings (both flavour conserving and flavour 
changing ones) that appear either in eq.~(\ref{kg}) or in its simplified 
versions, eqs.~(\ref{kg1}) and (\ref{kg2}).
In analogy to our discussion on $\de \rho$, we can see that 
bounds are not very restrictive at present, but will be important 
when  $\kappa_g$ and $\kappa_{\ga}$ are measured more precisely,
first at the LHC and then at future colliders. Thus the
processes $gg \leftrightarrow h$ and $h\to \ga \ga$
can be sensitive probes not only of
stop parameters \cite{dj98,djou,stops,hlm}, but also, more generally,
of the full stop/scharm sector\footnote{We also recall that
even direct searches for stops are affected by the presence
of $\tca$ and/or $\cta$. For instance, since these parameters induce 
stop/scharm mass mixing, decays such as 
$\st_i \to c {\tilde\chi}^0$ can proceed at the tree level.}.


\section{The decay $h \to c \, \ov{c}$}

In the previous Section we have discussed Higgs couplings to
gluons or photons, where the leading SM amplitudes 
arise at the one-loop level. SUSY corrections are
potentially important because they contribute
at the same perturbative order and are only
suppressed by the usual decoupling factor $v^2/\mt^2$, 
relatively to the SM.
The case of Higgs couplings to fermions has both similarities 
and differences with the previous one. In fact, in the SM,
Yukawa couplings are present at the tree level, but they
are suppressed for light fermions because they
are proportional to fermion masses. Therefore loop corrections
from new physics can be important if they do not respect that 
proportionality. 
In the special framework discussed in our paper, such a
situation arises for the charm quark provided all three
trilinears ($X_t$, $\tca$ and $\cta$) are simultaneously 
present and unsuppressed. 

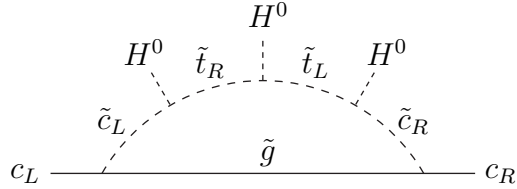
\begin{figure}[tb]
\begin{center}
\begin{picture}(300,60)(-150,0)
\Line(-80,0)(80,0)
\DashCArc(0,-35)(70,30,150){3}
\DashLine(-35,26)(-42,38){2}
\DashLine(35,26)(42,38){2}
\DashLine(0,35)(0,50){2}
\Text(0,8)[]{ $\glu$}
\Text(-90,0)[]{$c_L$}
\Text(90,0)[]{$c_R$}
\Text(-57,20)[]{$\scl$}
\Text(57,20)[]{$\scr$}
\Text(-20,42)[]{$\str$}
\Text(20,42)[]{$\stl$}
\Text(-45,46)[]{$H^0$}
\Text(47,46)[]{$H^0$}
\Text(2,60)[]{$H^0$}
\end{picture}
\end{center}
\vspace{-0.5cm}
\caption{\em 
One-loop stop/scharm/gluino contribution to the
effective Higgs coupling to charm quarks.} 
\label{ccfig}
\end{figure}

Before presenting our computation, let us recall again
the phenomenological $\kappa_i$ parametrization \cite{LHCwg},
which in the case of the decay $h \to c {\ov c}$ reads as
\be
\label{gammacc}
\frac{\Gamma(h\to c {\ov c})}{\Gamma^{\rm SM}(h\to c {\ov c})} =
\kappa_c^2 \simeq 1 + 2 \, \de \kappa_c \; ,
\ee
where we have expanded $\kappa_c=1+\de \kappa_c$ and 
$\de \kappa_c={\cal O}(v^2/\mt^2)$ encodes the corrections from new physics.
At the lagrangian level, using the previous expression amounts
to parametrize the mass of the charm quark and its effective coupling 
to the physical Higgs boson $h$ as
\be
\label{lcch}
{\cal L}_{c} = - m_c \left( 1 + \kappa_c \frac{h}{\sqrt{2}v} \right)
\ov{c} c  \; .
\ee 
In terms of the Higgs field $H^0$, the SM limit ($\kappa_c=1$)
is described by the $d=4$ Yukawa operator 
$-(y_c H^0 \ov{c_R} c_L + {\rm h.c.})$, whereas
the leading effects of non-SM physics are
associated with the $d=6$ effective operator 
$C_c |H^0|^2 H^0 \ov{c_R} c_L + {\rm h.c.}$ \cite{bw},
which we treat as a small perturbation\footnote{This
is justified {\em a posteriori}. Also, we consider a field
basis in which $y_c$ is real and neglect the effect of 
${\rm Im}\, C_c$, which induces a CP violating coupling
of $h$ to $i{\ov c} \gamma_5 c$ and corrects the ratio in
eq.~(\ref{gammacc}) at ${\cal O}(v^4/\mt^4)$.}.
The latter contributes to the charm mass and to the Higgs-charm 
coupling with different numerical coefficients,
{\it i.e.} $\de m_c = -({\rm Re}\, C_c)\, v^3$ and
$\de(\kappa_c m_c) = -3 \, ({\rm Re} \, C_c) \, v^3$,
so the physically relevant correction is 
$\de\kappa_c = -2 \, ({\rm Re} \, C_c) \, v^3/m_c$.
In our scenario, the dominant contribution to that $d=6$ effective 
operator is generated by the one-loop stop/scharm/gluino diagram 
shown in Fig.~\ref{ccfig}.
%
%
A crucial feature of such a diagram is that the chiral transition
from $c_L$ to $c_R$ does not involve the charm Yukawa coupling,
since it occurs through three chirality flips
associated with the trilinear couplings $X_t,\tca,\cta$.
We obtain:
\be
\label{dekac}
\de\kappa_c = - \frac{4\alpha_s}{3\pi} 
\left( \frac{m_t}{m_c} \right) 
m_t^2 \, {\rm Re} (\cta X_t^* \tca  M_g^*) \cdot 
I(|M_g|^2,\mmcl,\mmcr,\mmtl,\mmtr) \; ,
\ee
where $M_g$ is the gluino mass and $I$ is the five-point loop function:
\be
\label{i5}
I(a,b,c,d,e)= \frac{b \log(b/a)}{(b-a)(b-c)(b-d)(b-e)} +
(b \leftrightarrow c)+(b \leftrightarrow d)+(b \leftrightarrow e)
\; .
\ee
The simplified scenario in which $\mmtl \simeq \mmtr$ and 
$\mmcl \simeq \mmcr $ is described by the following
limit of eq.~(\ref{i5}):
\be
\label{i5s}
I(a,b,b,c,c)=\frac{1}{a(b-c)^2} \left[
 f_1 \! \left(\frac{b}{a}\right)+
 f_1 \! \left(\frac{c}{a}\right)
- \frac{2a}{b-c}
\left(\frac{b \log(b/a)}{b-a}-\frac{c \log(c/a)}{c-a}\right) 
\right] \; ,
\ee
where $f_1(x)$ has been defined in eq.~(\ref{f1f2}).
Eq.~(\ref{i5s}) can also be used in a scenario with 
$\mmtl \simeq \mmcl$ and $\mmtr \simeq \mmcr$, because 
$I(a,b,c,b,c)=I(a,b,b,c,c)$.
Finally, in the degenerate limit in which all stop and scharm squarks 
have a common mass $\mt^2$, the result reads:
\be
\label{dekacs}
\de\kappa_c = - \frac{4\alpha_s}{3\pi} 
\left( \frac{m_t}{m_c} \right) 
\left( \frac{m_t^2}{\mt^2} \right) \left[
\frac{ {\rm Re}(\cta X_t^* \tca M_g^*)}{\mt^4}  
\, g_1 \! \left(\frac{|M_g|^2}{\mt^2}\right) \right] \; ,
\ee
where $g_1(x)$ has been defined in eq.~(\ref{g1g2}) [note
that $g_1(|M_g|^2/\mt^2) \to 1/12$ when  $|M_g|^2 \to {\mt^2}$].

The main properties of the correction $\de\kappa_c$ are 
manifest both in our general result, eq.~(\ref{dekac}),
and in its simplified version, eq.~(\ref{dekacs}).
Consider the latter expression.
One can notice that the loop factor and the decoupling factor 
$m_t^2/\mt^2$, which suppress $\de\kappa_c$, are partly
compensated by the large enhancement factor $m_t/m_c$, whose value 
is about $2.7 \cdot 10^2$ at the weak or SUSY scale. 
The expression in square brackets
is a dimensionless function of SUSY mass parameters. The general
case is described by eq.~(\ref{dekac}). As a numerical example,
suppose that the gluino, stop and scharm masses as well as the three 
trilinear parameters have a common size of about 1~TeV. Then  
$\de\kappa_c \sim \pm 3 \cdot 10^{-2}$, which implies a $6 \%$
deviation of $\Gamma(h\to c {\ov c})$ with respect
to the SM prediction. However, it is clear that even moderate 
variations around that parameter point can generate very different 
results. For instance, increasing all trilinear
parameters to 1.5~TeV while keeping squark and gluino masses
at 1~TeV would lead to a sizeable $20 \%$ deviation in
$\Gamma(h\to c {\ov c})$,
while a similar change with reversed roles would reduce the 
deviation to $1 \%$ only. Splitting squark masses
can produce further variations\footnote{The above corrections 
to the Higgs-charm effective coupling could be compared with 
different ones, which arise from integrating out the heavy Higgs 
doublet. In the MSSM, for instance, one obtains a tree-level
contribution to the $|H^0|^2 H^0 \ov{c_R} c_L$ effective operator such that 
$\de\kappa_c \simeq 2 \cos 2\beta \cos^2\! \beta \cdot m_Z^2/m_A^2$,
where $m_A$ is the heavy Higgs mass. This correction is negative
and its size is smaller than $10^{-2}$ for $m_A \simgt 500$~GeV.}.

As a final comment, we recall that the sensitivity
of $\Gamma(h\to c {\ov c})$ to flavour violation 
in the stop/scharm sector of the MSSM has been recently 
pointed out and explored in \cite{beghm}. Flavour violating terms
of all types (LL,RR,LR) have been considered and
potentially large effects have been reported. 
In our study we have focused
on LR flavour violation, used a simpler computational
method and obtained simpler analytical expressions,
which expose the relevant parametric dependences.
Although a quantitative comparison between our 
results and those in \cite{beghm} is not 
straightforward, we have noticed that the effects reported there 
are typically larger than those we find, even if we extend our
approach to include LL and RR violation. We do not
have an explanation for such a discrepancy.
At the qualitative level, though, we confirm the potential
relevance of SUSY flavour violation to the decay 
$h\to c {\ov c}$ and agree with the conclusion 
in \cite{beghm} that such effects may be tested 
at a future $e^+ e^-$ collider through precision
measurements, 
while that task will be hard at the LHC 
because of the difficulties in charm tagging\footnote{
For recent studies on the observability of the Higgs-charm
coupling, see also \cite{psst} and refs. therein. 
Regarding other interesting processes
that involve the Higgs boson and charm quarks, 
we recall that $\tca$ or $\cta$ can also induce the 
decay $t \to c h$ through one-loop diagrams. 
According to \cite{tch}, however, the associated 
branching ratio can hardly exceed ${\cal O}(10^{-6})$, which is
below the LHC sensitivity.}. 


\section{Conclusions}

After the discovery of the Higgs boson at the LHC, many efforts will
be devoted to measure its couplings more precisely and to
look for possible deviations from the SM expectations.
At the same time, the search for new particles will continue
and higher mass ranges will be probed.
The framework beyond the SM that we have examined in this paper
is a SUSY scenario with large flavour 
violating $A$-terms in the stop/scharm sector.
By integrating out stop and scharm squarks at the one-loop level,
we have computed the leading corrections induced by $\tca$ and $\cta$
on the Higgs mass, the electroweak $\rho$ parameter and 
the effective Higgs couplings to gluons, photons and charm quarks.
For each of such quantities we have presented explicit analytical 
expressions which exhibit the relevant parametric dependences, 
both in the general case and in special limits.
In particular, by treating $\tca$ and $\cta$ on the 
same footing as the flavour conserving parameter $X_t$,
we have emphasized that all three trilinear couplings play similar 
roles and can induce significant effects. 
We have also checked that each of the above observables has the 
correct scaling behaviour under the decoupling of SUSY particles,
as expected from the dimensionality ($d=4$ or $d=6$) of the 
associated effective operators, and have discussed some
phenomenological implications at the LHC and future colliders.
It is also clear that the importance of the indirect SUSY
effects investigated in this paper is both related
and complementary to the results of ongoing direct searches of
SUSY particles.

\vspace{1 cm}

{\bf Acknowledgements}: We thank P.~Paradisi for discussions.



\end{document}